\documentclass[twocolumn,tighten,times]{aastex62}

\usepackage{amssymb,amsmath}
\usepackage{graphicx}
\usepackage{color}
\usepackage{subfigure}
\usepackage{lineno}

\newcommand{\mgii}{\ion{Mg}{2}}

\newcommand{\iris}{{\em IRIS}}

\submitjournal{ApJL}

\shorttitle{}
\shortauthors{Mart\'inez-Sykora et al.}

\newcommand{\longacknowledgment}{We gratefully acknowledge support by NASA grants 80NSSC20K1272, 80NSSC21K0737, 80NSSC21K1684, and contract NNG09FA40C (\iris). Resources supporting this work were provided by the NASA High-End Computing (HEC) Program through the NASA Advanced Supercomputing (NAS) Division at Ames Research Center. The simulations have been run on the Pleiades cluster through the computing project s1061, and s2601. This research is also supported by the Research Council of Norway through its Centres of Excellence scheme, project number 262622, and through grants of computing time from the Programme for Supercomputing. Data are courtesy of \iris. \iris\ is a NASA small explorer mission developed and operated by LMSAL with mission operations executed at NASA Ames Research Center and major contributions to downlink communications funded by ESA and the Norwegian Space Centre.The radiative transfer computations were enabled by resources provided by the Swedish National Infrastructure for Computing (SNIC) at the PDC Center for High Performance Computing, KTH Royal Institute of Technology, partially funded by the Swedish Research Council through grant agreement no. 2018-05973. JdlCR gratefully acknowledges financial support from the European Research Council (ERC) under the European Union's Horizon 2020 research and innovation program (SUNMAG, grant agreement 759548).}

\begin{document}

\title{Chromospheric Heating from Local Magnetic Growth and Ambipolar Diffusion Under Non-Equilibrium Conditions}

\correspondingauthor{Juan Mart\'inez-Sykora}
\email{martinezsykora@baeri.org}

\author[0000-0002-0333-5717]{Juan Mart\'inez-Sykora}
\affil{Lockheed Martin Solar \& Astrophysics Laboratory, 3251 Hanover Street, Palo Alto, CA 94304, USA}
\affil{Bay Area Environmental Research Institute, NASA Research Park, Moffett Field, CA 94035, USA}
\affil{Rosseland Centre for Solar Physics, University of Oslo, P.O. Box 1029 Blindern, NO-0315 Oslo, Norway}
\affil{Institute of Theoretical Astrophysics, University of Oslo, P.O. Box 1029 Blindern, NO-0315 Oslo, Norway}

\author[0000-0002-4640-5658]{Jaime de la Cruz Rodr\'iguez}
\affil{Institute for Solar Physics, Dept. of Astronomy, Stockholm University, AlbaNova University Centre, SE-10691 Stockholm, Sweden}

\author[0000-0002-5879-4371]{Milan Go\v{s}i\'{c}}
\affil{Lockheed Martin Solar \& Astrophysics Laboratory, 3251 Hanover Street, Palo Alto, CA 94304, USA}
\affil{Bay Area Environmental Research Institute, NASA Research Park, Moffett Field, CA 94035, USA}

\author[0000-0002-3234-3070]{Alberto Sainz Dalda}
\affil{Lockheed Martin Solar \& Astrophysics Laboratory, 3251 Hanover Street, Palo Alto, CA 94304, USA}
\affil{Bay Area Environmental Research Institute, NASA Research Park, Moffett Field, CA 94035, USA}

\author[0000-0003-0975-6659]{Viggo H. Hansteen}
\affil{Lockheed Martin Solar \& Astrophysics Laboratory, 3251 Hanover Street, Palo Alto, CA 94304, USA}
\affil{Bay Area Environmental Research Institute, NASA Research Park, Moffett Field, CA 94035, USA}
\affil{Rosseland Centre for Solar Physics, University of Oslo, P.O. Box 1029 Blindern, NO-0315 Oslo, Norway}
\affil{Institute of Theoretical Astrophysics, University of Oslo, P.O. Box 1029 Blindern, NO-0315 Oslo, Norway}

\author[0000-0002-8370-952X]{Bart De Pontieu}
\affil{Lockheed Martin Solar \& Astrophysics Laboratory, 3251 Hanover Street, Palo Alto, CA 94304, USA}
\affil{Rosseland Centre for Solar Physics, University of Oslo, P.O. Box 1029 Blindern, NO-0315 Oslo, Norway}
\affil{Institute of Theoretical Astrophysics, University of Oslo, P.O. Box 1029 Blindern, NO-0315 Oslo, Norway}

\begin{abstract}

The heating of the chromosphere in internetwork regions remains one of the foremost open questions in solar physics. In the present study we tackle this old problem by using a very high spatial-resolution simulation of quiet-Sun conditions performed with radiative MHD numerical models and \iris\ observations. We have expanded a previously existing 3D radiative MHD numerical model of the solar atmosphere, which included self-consistently locally driven magnetic amplification in the chromosphere, by adding ambipolar diffusion and time-dependent non-equilibrium hydrogen ionization to the model. The energy of the magnetic field is dissipated in the upper chromosphere, providing a large temperature increase due to ambipolar diffusion and the non-equilibrium ionization (NEQI). At the same time, we find that adding the ambipolar diffusion and NEQI in the simulation has a minor impact on the local growth of the magnetic field in the lower chromosphere and its dynamics. Our comparison between synthesized \mgii\ profiles from these high spatial resolution models, with and without ambipolar diffusion and NEQI, and quiet Sun and coronal hole observations from \iris\ now reveal a better correspondence. The intensity of profiles is increased and the line cores are slightly broader when ambipolar diffusion and NEQI effects are included. Therefore, the \mgii\ profiles are closer to those observed than in previous models, but some differences still remain.
\end{abstract}


\section{Introduction}

In the chromosphere, a complex force balance originating from gas pressure gradients and magnetic forces gives rise to a highly dynamic and complex atmospheric structure given the ubiquitous presence of non-linear waves \citep[e.g.,][]{Carlsson+Stein1994,Carlsson:2002wl} and small-scale magnetic fields \citep[e.g.,][]{Lites:2008ss}. 
However, comparisons between radiative (M)HD models and high-resolution observations showed insufficient wave power in the upper chromosphere \citep[e.g.,][]{Carlsson:2007uq}, although this remains a topic of discussion \citep{Molnar2022PhDT........13M}. 
On the other hand, the weak magnetic fields that continuously emerge on granular scales, and that are thought to be generated by a local dynamo, have also been detected in numerical models \citep{abbett2007} and they have enormous potential for heating the low solar atmosphere. Recently, \citet{Amari:2015fe} used to models to suggest that the local dynamo in the upper convection zone could maintain the heating of the whole solar atmosphere. However, radiative MHD models indicate that the strong entropy drop at the solar surface leads to a net negative Poynting flux and strong super-adiabaticity, making it very difficult for the magnetic field to reach the chromosphere \citep[e.g.,][]{abbett2007,Nordlund:2008dq,Moreno-Insertis:2018hl}. Similarly, there has been limited observational evidence for tracking internetwork emerging fields from the photosphere into the chromosphere and above, and the production of localized heating \citep[e.g.,][]{Martinez-Gonzalez:2009rp,Gosic:2016vk,Gosic:2021ApJ...911...41G}. 

In addition to waves and shocks, another potential mechanism to transfer energy to greater heights are the swirls \citep[e.g.,][]{Yadav:2020ApJ...894L..17Y}. However, in principle, this mechanism does not lead to the transport of the magnetic flux but kinetic energy and its role in heating the chromosphere is unclear. As an alternative scenario, \citet{Martinez-Sykora:2019dyn} showed that the magneto-acoustic shocks in the chromospheric internetwork (IN) regions in quiet sun (QS) or coronal hole (CH) could convert kinetic energy into magnetic energy, i.e., producing a local magnetic energy growth in the chromosphere. That study lacked a proper model of ambipolar diffusion and NEQI for hydrogen. The former represents an efficient mechanism to dissipate currents~\citet{Khomenko:2012bh,Martinez-Sykora:2017gol}, whereas the latter has a profound effect on the temperature, ionization fraction and electron densities in the chromosphere \citep{Leenaarts:2007sf,Golding:2016wq}. Knowing the ionization fraction accurately is necessary in order to calculate the ambipolar diffusivity and therefore including NEQI of hydrogen is required in order to also model precisely the heating caused by ambipolar diffusion~\citep{Khomenko:2014nr,Martinez-Sykora:2020ApJ...889...95M,Nobreg-Siverio:2020AA...633A..66N}. In the current work, we expand the model presented in \citet{Martinez-Sykora:2019dyn} by including ambipolar diffusion, and the hydrogen ionization is treated under NEQI conditions, allowing us to investigate its role in chromospheric heating.

\section{Numerical models}~\label{sec:mod}

Our 3D radiative-MHD numerical simulation is computed with the Bifrost code \citep{Gudiksen:2011qy}. The model includes radiative transfer with a scattering in the photosphere and lower chromosphere \citep{Skartlien2000,Hayek:2010ac}. In the middle and upper chromosphere, radiation from specific species such as hydrogen, calcium, and magnesium is computed following \citet{Carlsson:2012uq} recipes while using optically thin radiative losses in the transition region and corona. Thermal conduction along the magnetic field is important for the energetics of the transition region and corona. The main difference with the model presented in \citet{Martinez-Sykora:2019dyn} is that the current simulation treats the hydrogen ionization in non-equilibrium conditions \citep{Leenaarts:2007sf} and the ion-neutral interaction effects by including the ambipolar diffusion \citep{Martinez-Sykora:2020ApJ...889...95M,Nobreg-Siverio:2020AA...633A..66N}. 

The simulation spans in the vertical axis a range of heights from $\sim2.5$~Mm below the photosphere to $8$~Mm above into the corona with a non-uniform vertical grid size with the smallest grid size of $4$~km in the photosphere and chromosphere. The photosphere is located at $z=0$, where $\tau_{500} \sim 1$. The horizontal domain spans $6\times6$~Mm in the $x$ and $y$ directions with $5$~km resolution. Initially, the simulation box is seeded with a uniform weak vertical magnetic field of $2.5$~G. The convective motion builds magnetic field complexity, and as result the magnetic field strength reaches a statistical steady-state with $B_{rms} = 57$~G, average $|B_z|=17$~G with some flux concentrations reaching 2kG at photospheric heights \citep[similar to that described by][]{Vogler:2007yg,Rempel:2014sf,Cameron:2015rm}. 


In this high-resolution quiet Sun model, the chromospheric magnetic energy content increases substantially with time, especially in the region that is dominated by shock waves and where plasma $\beta > 1$ ($0.5<z<2$~Mm). Essentially, the magnetic energy is growing in place while it is being fed by the dynamics of the model. The magnetic growth is produced by colliding shocks and shear flows instead of the turbulent motion of the convective cells in the convection zone. Further details of this model can be found in \citet{Martinez-Sykora:2019dyn}. After the magnetic field has reached statistical steady state, some $\sim25$ minutes into the run, ambipolar diffusion and NEQI are turned on. The simulation is then run for another $\sim10$ minutes, which allows the transition from LTE to a NEQI and ambipolar diffusion to settle. 

\section{Spectral synthesis}~\label{sec:syn}

We have computed the \mgii\ synthetic profiles using the 1.5D parallel version of the RH radiative transfer code \citep{Uitenbroek:2001dq,Pereira:2015th}. The model atom consists of 10 bound levels plus the \ion{Mg}{3} continuum \citep{Leenaarts:2013ij}. Partial redistribution of the frequency in the scattered radiation (PRD) has been included in the modelling of the \mgii\ h\&k lines by using the fast angle approximation proposed by \citep{Leenaarts2012AA...543A.109L}. The radiative transfer equation was solved using cubic Bezier splines solvers \citep{delaCruzRodriguez:2013ApJ...764...33D}.

\section{Observations}~\label{sec:obs}

In this manuscript we used the already analyzed \iris\ \citep{De-Pontieu:2014yu} dataset  presented in  \citet{Martinez-Sykora2022arXiv221015150M}. The observations were obtained on 2016 March 25 and 2017 October 15. The observations include an equatorial coronal hole (CH), starting at 10:09:18~UT and ending 11:58:45~UT. The observed a quiet Sun region from 17:54:22~UT until 22:52:35~UT. Both observations were taken at disk center. We separate the internetwork field (IN) from the network field (NE), and refer to \citet{Martinez-Sykora2022arXiv221015150M} for further details. 

\section{Results}~\label{sec:res}

\begin{figure*}
    \centering
    \includegraphics[width=0.98\textwidth]{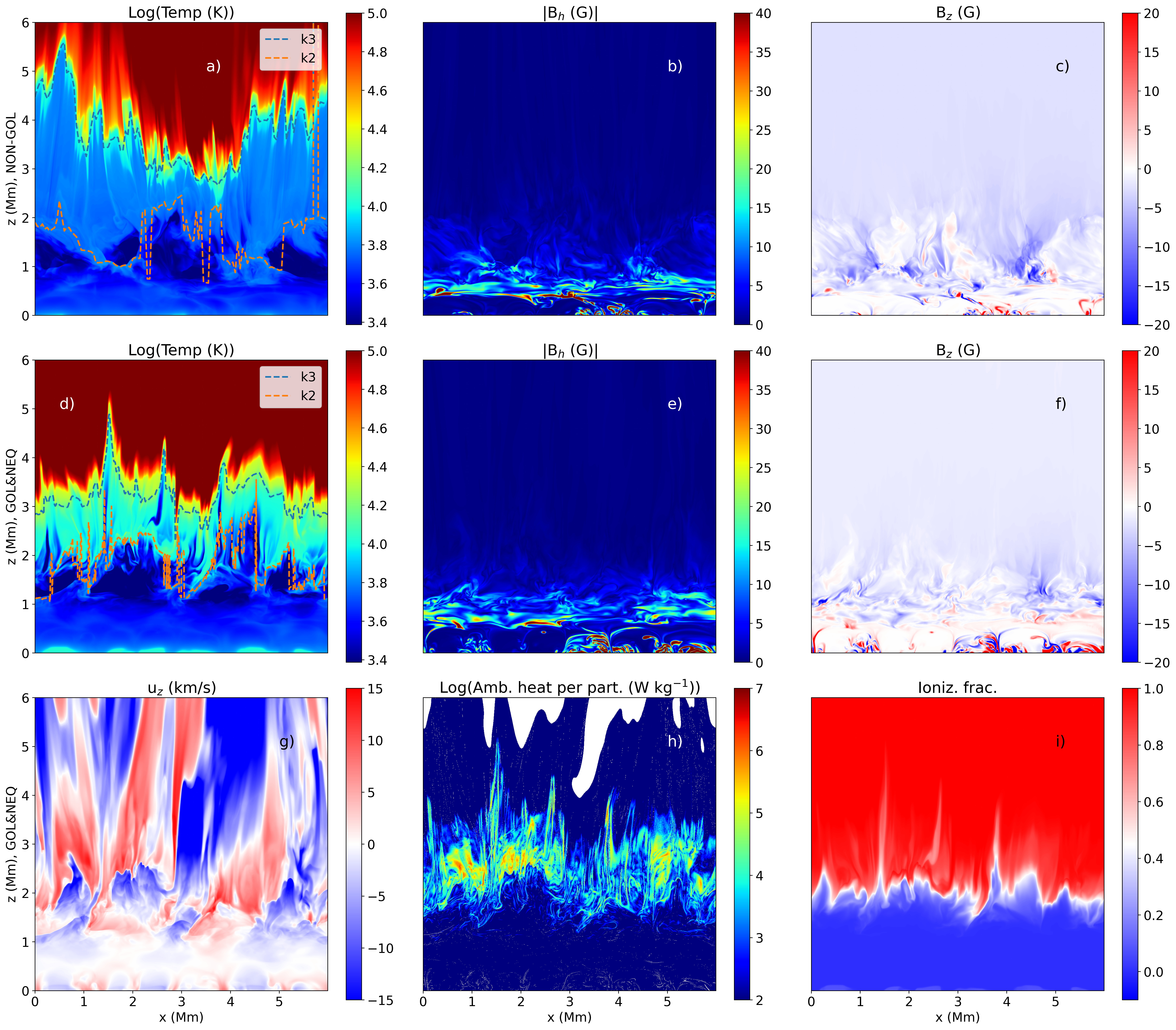}
    \caption{Top row shows the simulation without NEQI and ambipolar diffusion (non-GOL) and second and third row includes NEQI and ambipolar diffusion (GOL\&NEQ). Panels a to c and  d to f show vertical cut maps of temperature, horizontal field strength, and vertical field strength, respectively. For GOL\&NEQ, panels g to i show vertical velocity, ambipolar heating, and ionization fraction. The height for optical depth ($\tau$) equal to unity at the \mgii\ $k_2$ and $k_3$ locations has been added in panel a and d. In the convection zone ($z<0$), the magnetic field is accumulated in the downflows. Shocks dominate the simulated chromosphere ($0.8<z<2$~Mm) and jets ($2<z<4$~Mm), whereas the corona is maintained by energy release from magnetic braiding and dominated by thermal conduction.}
    \label{fig:vercut}
\end{figure*}

In both simulations, i.e., without and with ambipolar diffusion and NEQI the vertical magnetic field slice (panels c and f in Figure~\ref{fig:vercut})  reveals the highly complex salt and pepper mixture in the low and middle chromosphere ($z\sim[0.5,2]$~Mm) induced by the kinematics as described in detail by \citet{Martinez-Sykora:2019dyn} and \citet{Martinez-Sykora2022arXiv221015150M}. This magnetic energy is created in situ due to the shocks (panel b) traveling through, which fold,  stretch, twist, and reconnect the magnetic field, thereby building up its energy. 

\begin{figure}
    \includegraphics[width=0.49\textwidth]{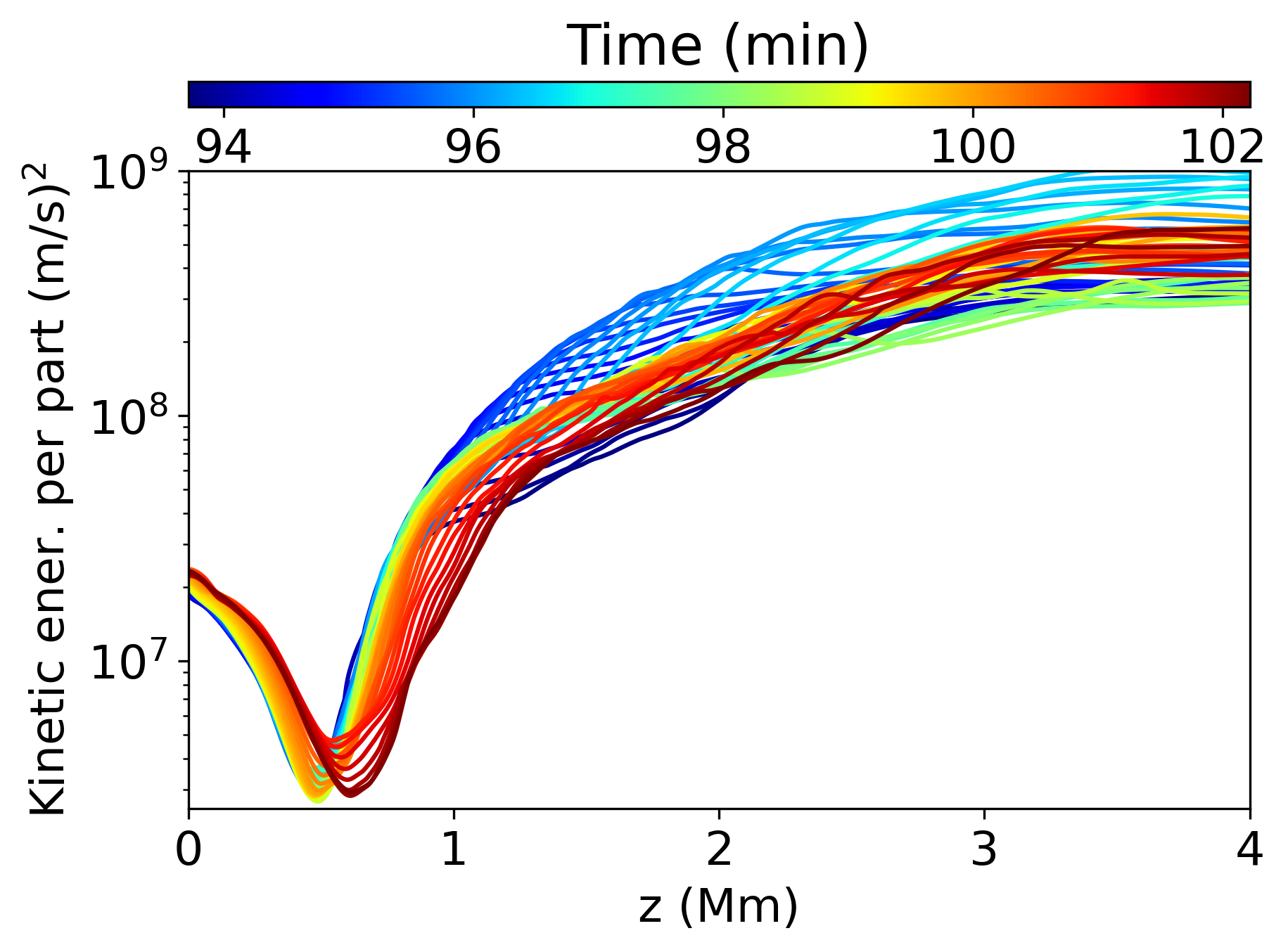}
    \includegraphics[width=0.49\textwidth]{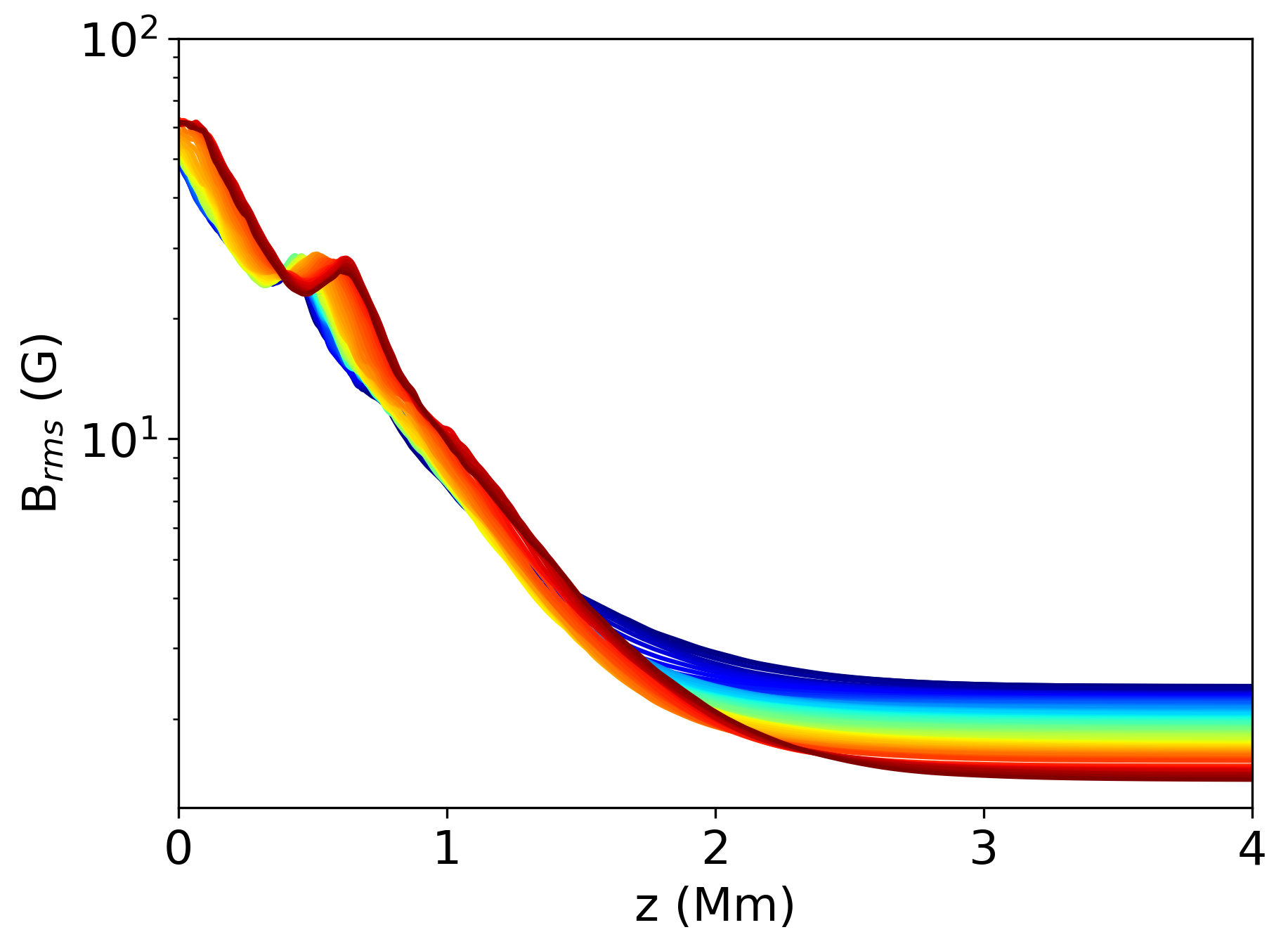}
    \caption{The magnetic field strength and kinetic energy per particle r.m.s as a function of height (horizontal axis) and time (colorbar) reveals that the ambipolar diffusion and NEQI effects do not change substantially the magnetic field growth of kinetic energy. The time range in this plot showed the evolution after both effects have been turned on. The ambipolar diffusion dissipates magnetic energy in the upper chromosphere.}
    \label{fig:rms}
\end{figure}

The kinetic energy per particle does not change significantly with time as the ambipolar diffusion and NEQI effects are turned (top panel of Figure~\ref{fig:rms}). We see only that there is a small shift to greater heights (by $\sim100$~km) of the profile with height. 
 
Interestingly, the ambipolar diffusion (and ambipolar heating, panel h in Figure~\ref{fig:vercut}) is taking place just at the top (upper-chromosphere, $z\sim[1.5,3.5]$~Mm) of the location of the in-situ chromospheric magnetic growth (lower- and mid-chromosphere, $z\sim[0.5,2]$~Mm). Consequently, the ambipolar diffusion is not ``annihilating" the local magnetic growth due to the conversion from kinetic to magnetic energy.  This can be appreciated in the time evolution of the magnetic field strength and kinetic energy per particle r.m.s as a function of height once the NEQI and ion-neutral interaction effects have been turned on. The bottom panel of Figure~\ref{fig:rms} shows that the magnetic field is maintained in time in the low and mid chromosphere and dissipated in time at greater heights.


The ambipolar diffusion dissipates substantial magnetic energy that is introduced into the mid and upper chromosphere from in-situ chromospheric magnetic growth. Consequently, the magnetic energy in the upper chromosphere and above decreases with time (bottom panel of Figure~\ref{fig:rms}). The magnetic energy in the upper chromosphere arises from the complex evolution of the magnetic field due to  waves, currents, and colliding shocks traveling through the atmosphere. It is these currents that are dissipated by ambipolar diffusion. 

The effects of the dissipation of these currents has a significant effect on the thermodynamics of the upper chromosphere. These effects are substantially modified by NEQI effects. Compared to LTE ionization, the spatial maps of ionization degree in NEQI show a less sharp transition from partial ionization to full ionization (panel i in Figure~\ref{fig:vercut}). In addition, it is known that recombination takes longer so the plasma remains ionized for longer  \citep[][]{Leenaarts:2007sf,Martinez-Sykora:2020ApJ...889...95M}. The latter two papers showed that under NEQI conditions, dissipation of energy causes a temperature increase for a short time (of order several seconds, depending on the local conditions) before ionization occurs, instead of immediately ionizing the plasma. So we see in our NEQI simulation that the dissipation of currents (through ambipolar diffusion) raises the temperature by several thousands of Kelvin, which produces an upper chromosphere of $\sim10^4$~K, as seen by comparing panels a and d from Figure~\ref{fig:vercut}.  Note that the location of the strongest magnetic field values or of the velocities in the chromosphere ($z=[0.5,1.5]$~Mm) has, statistically speaking, barely changed at that instance (compare panels b and c with e and f). 

\begin{figure*}
    \centering
    \includegraphics[width=0.94\textwidth]{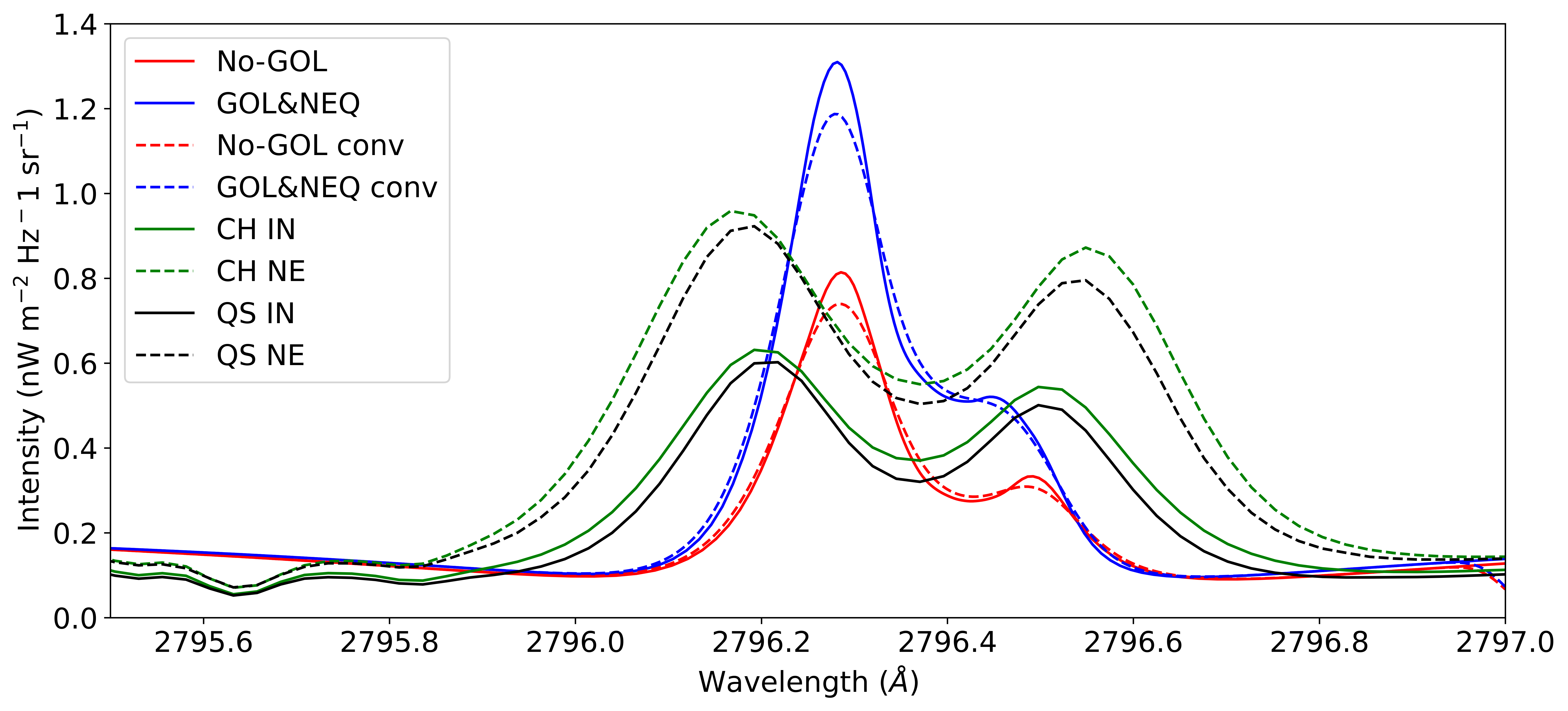}
    \caption{The synthesis of  \mgii~k before (red) and after (blue) the ambipolar diffusion and NEQI has been turned on are compared with QS (black) and CH (green) \iris\ observations. The observations have been separated into IN (solid) and NE (dashed) field regions, as detailed in \citet{Martinez-Sykora2022arXiv221015150M}. In addition, the synthetic profiles have been convolved with \iris\ instrumental spectral broadening ($\sim 0.056$~\AA\ FWHM, dashed lines).}
    \label{fig:syn_obs}
\end{figure*}

We have computed the radiative transfer for \mgii\ chromospheric lines for these two instances shown in Figure~\ref{fig:vercut}. However, before we compare the synthesis of these two models and with \iris\ observations, we point out that $\tau =1$ within the wavelength range of $k_2$ and $k_3$ is within the upper chromosphere in both models (contours in panel a and d of Figure~\ref{fig:vercut}). Consequently, the line core captures the region that has been heated due to local magnetic growth and ambipolar diffusion. 

We compare synthetic \mgii~k line from on-disk QS and CH targets separating IN and NE fields. Due to the highly simplified and weak initial magnetic field and dimensions of the numerical domain, the model is closer to mimicking an IN region in a coronal hole or quiet sun than a NE field. The synthesis before (red) and after (blue) of the ambipolar diffusion and NEQI are turned on are compared with the observations (Figure~\ref{fig:syn_obs}). In addition, we consider \iris\ spectral broadening by convolving the synthetic profiles with a Gaussian with an FWHM of 0.056~\AA\ \citep[dashed lines,][]{Pereira:2013ys}. 

We find that the synthetic profiles (solid lines) are asymmetric with a very strong blue peak. This asymmetry comes from the fact that $k_3$ is red-shifted, which can be caused by a non-zero net downflow in the upper chromosphere. A more complex field topology with NE surrounding the IN field may produce a balanced net flow in the upper chromosphere and center the \mgii~$h_3$/$k_3$ wavelength location more. 

The convolution also broadens the profiles slightly and reduces the peak intensity. The NEQI and ambipolar effects lead to profiles with stronger intensities, and the profiles are also somewhat broader than without those effects. \citet{Carlsson:2015fk} showed that a hotter and denser upper chromosphere can broaden \mgii\ profiles. Nevertheless, despite the combination of including effects from ambipolar diffusion, local magnetic growth, and NEQI effects that result in a hotter chromosphere, and the \iris\ instrumental broadening, the spectral profiles remain slightly narrower than in the observations. This suggests that this model may be missing more physical processes or includes a too simplified magnetic field topology; see discussion for further details. 

\section{Discussion}~\label{res:dis}

\citet{Martinez-Sykora:2019dyn} presented a new possible mechanism for generating chromospheric magnetic field by converting the kinetic energy into magnetic energy in situ in the quiet Sun or coronal hole internetwork regions. In this work, we have extended this model by now adding NEQI of hydrogen and ambipolar diffusion. We found that the magnetic energy growth in the chromosphere continues to occur in the presence of these two physical processes. The ambipolar diffusion becomes important just above the layer in which the magnetic field growth occurs in the chromosphere. Ambipolar diffusion helps to convert magnetic energy into thermal energy in the upper chromosphere. The NEQI effects ensure that the ionization fraction is such that the ambipolar diffusion dissipates magnetic energy in the upper chromosphere and that this dissipation leads to an increase in temperature instead of ionizing the plasma. As a result, the upper chromosphere reaches temperatures above $\sim 10^4$~K. Note that NEQI effects are crucial to accurately estimate the ambipolar diffusion since the ambipolar diffusivity is proportional to the ionization degree \citep[see also][for different topologies]{Martinez-Sykora:2020ApJ...889...95M,Nobreg-Siverio:2020AA...633A..66N}.

The hotter upper chromosphere that we find in the new model has an effect on the spectral diagnostics that are formed this region.
We have compared the \iris\ \mgii\ profiles observations of CH and QS regions and separate between locations of IN and NE fields, similar to what was done in \citet{Martinez-Sykora2022arXiv221015150M}. The synthetic \mgii\ peaks from our models are too skewed towards the blue compared to the observations. We note that the field configuration in our simulation is highly simplified without any NE field surrounding the IN field and a relatively weak coronal magnetic field. This simplified field topology may contribute to having larger net flows in the upper chromosphere and skew the \mgii\ profiles. 

In addition, we have found that the profiles increase in intensity and slightly broaden due to a temperature increase in the upper chromosphere when the NEQI effects and ambipolar diffusion are considered. We find that observations show somewhat broader profiles than the synthetic observables. Therefore, the radiative MHD model or the radiative transfer treatment for the synthesis may be missing some physical processes. On the one hand, the radiative transfer calculations are based on the 1.5D approximation ignoring 3D radiative transfer effects, which may play a role in the core of the \mgii\ line \citep{Judge2020ApJ...901...32J}. It is unclear however how these 3D effects would lead to broader profiles. On the other hand, the narrower \mgii\ core profile in the simulations could come from a lack of complexity in the LOS velocity field, turbulence, mass loading into the upper chromosphere, and/or a lack of heating. Extending the upper chromospheric conditions to deeper layers can also broaden the \mgii\ profiles. In our simulation, we assumed that the plasma could be treated as a single fluid. It is possible that micro-physics or other multi-fluid effects may lead to significant changes in turbulence, heating, and spatial extent of the region with upper chromospheric conditions, and thus to broader \ion{Mg}{2} profiles. For instance, \citet{Oppenheim2020ApJ...891L...9O,Evans2022arxiv.2211.03644} suggest that the colder regions of the simulated atmosphere is where the thermal Farley Buneman instability \citep{Dimant2022arXiv221105264D} will grow and, hence, the large-scale single fluid assumptions fail in those locations. Indeed, the Farley-Buneman and thermal instabilities could lead to micro-turbulence and heating \citep{Oppenheim2020ApJ...891L...9O,Evans2022arxiv.2211.03644}. 

\acknowledgements{\longacknowledgment} 

\bibliographystyle{aasjournal}
\bibliography{collectionbib.bib}

\end{document}